\documentclass[aps,prc,superscriptaddress,showpacs,twocolumn,amsmath]{revtex4}
\usepackage{graphicx}
\usepackage{multirow}
\usepackage{color,makecell}
\usepackage{float}
\begin{document}
\title{A Classical Interpretation of the Nonrelativistic Quark Potential Model: Color Charge Definition and the Meson Mass-Radius Relationship}
\author{ZhiGuang Tan}\thanks{tanzg@ccsu.edu.cn}\author {YouNeng Guo}\thanks{guoxuyan2007@163.com}
\author{ShengJie Wang}
\affiliation{School of Electronic Information and Electrical Engineering, Changsha University,Changsha,
410003,P.R.China}
\author{Hua Zheng}\thanks{zhengh@snnu.edu.cn}
\affiliation{School of Physics and Information Technology, Shaanxi Normal University, Xi'an, 710119, China}
\begin{abstract}
Quantum Chromodynamics (QCD) is the fundamental theory describing quark interactions, and various quark models based on QCD have been widely used to study the properties of hadrons, including their structures and mass spectra. However, unlike Quantum Electrodynamics (QED) and the Bohr model of hydrogen atom, there is no direct classical analogy for hadronic structures. This paper presents a classical interpretation of the nonrelativistic quark potential model, providing a more intuitive and visualizable description of strong interactions through the quantitative formulation of color charge and color flux. Furthermore, we establish the relationship between meson mass and its structural radius in the nonrelativistic framework and estimate the key parameters of our model using available data from $\eta_b(1S)$ and $\Upsilon(1S)$. We then extend this relationship to a broader range of excited meson states, obtaining their structural radii that show good agreement with the root mean square (RMS) radius or charge radius predicted by QCD calculations.
\end{abstract}

\keywords{Nonrelativistic Quark Potential Model; Color Charge; Color flux; Meson Structure; Mass spectra and radius}

\pacs{25.75.Dw,25.75.Gz}
\maketitle
\section{introduction}
When the Schr$\ddot{o}$dinger equation
\begin{equation}
\nabla^2\Psi+\frac{2m}{\hbar^2}[E-V(r)]\Psi=0 \label{sder}\end{equation}
is solved to obtain the eigenenergies and mass spectra of a meson system composed of a pair of positive and negative quarks, the potential function $V(r)$ of the system is pivotal, i.e., the so-called quark potential models \cite{pot1,pot2,pot3}. Among them, the Cornell potential \cite{corn}, proposed in the 1980s, has performed effectively. Most current potential models are based on it, incorporating various improvements or extensions \cite{corn1,corn2}. It is written as \cite{corn}
\begin{equation}
V(r)=-\frac{a}{r}+br,\label{poti}
\end{equation}
where $a$ and $b$ are two positive parameters. The first term in Eq. (\ref{poti}) is the Coulomb-like potential, while the second term is to take into account the property of  quark confinement. Thus, it is very difficult to separate a pair of attractive quarks. Solving Eq. (\ref{sder}) to obtain mass eigenstates and quantum properties of hadrons constitutes an approach to studying the nature of those hadrons \cite{corn3,corn4}. In order to obtain results close to experimental measurements, not only can the parameters be adjusted, but the expression for the potential function can also be extended. In Refs. \cite{corna,cornb}, the potential has been extended to a more general form
\begin{equation}
V(r)=ar^2 +br-\frac{c}{r}+\frac{d}{r^2}+e .\label{pota}
\end{equation}

However, except for the two terms as in Eq. (\ref{poti}), no physically reasonable explanations have been provided for the origins of the other terms in Eq. (\ref{pota}).

According to a classical interpretation, the quantum numbers that describe the properties of hadrons\cite{nst} are inherently linked to their internal structure. So the interaction between quarks should be determined by their color charge values, relative positions, motion states and spin orientations. This study aims to identify the physical origins for each term in the potential model Eq. (\ref{pota}) and provide a classical description of quark interactions.

The paper is organized as follows. In Sec. \ref{s2}, we introduce the concept of unit color charge and discuss the interaction between two stationary color charges in vacuum. We also provide the rule for the dot product of two color charges, along with the basic function, which corresponds to the Coulomb-like term in the potential Eq. (\ref{pota}). Section \ref{s3} explains the inverse square term in the potential by introducing the concepts of color flow and color magnetic field. Section \ref{s4} addresses the harmonic oscillator potential arising from spin, corresponding to the other three terms in Eq. (\ref{pota}). In Sec. \ref{s5}, we estimate the relevant model parameters and present our numerical results based on the classical description of meson structures, comparing them with data in the literature from potential models. Finally,  a brief summary and discussion are given in Sec. \ref{s6}.

\section{The interaction between a pair of stationary quarks in vacuum}\label{s2}
\begin{figure}[h!]
\center{\includegraphics[width=0.25\textwidth]{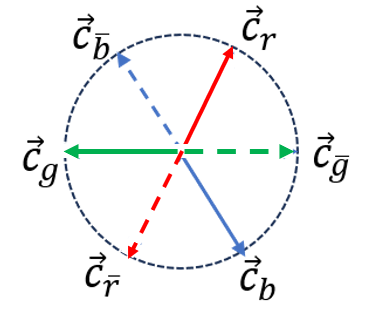}}
\caption{Vector representation of unit color charge.}\label{vecc}\end{figure}

 To provide a classical description of the interaction between a pair of quarks, we first define three fundamental color charges $c_r,c_g,c_b$ and their corresponding anti-color charges $c_{\bar{r}},c_{\bar{g}},c_{\bar{b}}$ as follows
\begin{eqnarray}
  &c_r \equiv e^{\theta i}   , \quad& c_{\bar{r}}\equiv e^{(\theta+\pi) i}=-c_r; \nonumber\\
  &c_g \equiv e^{(\theta+\frac{2\pi}{3})i},&c_{\bar{g}}\equiv e^{(\theta-\frac{\pi}{3})i}=-c_g;\label{crgb}\\
  &c_b \equiv e^{(\theta-\frac{2\pi}{3})i}, &c_{\bar{b}}\equiv e^{(\theta+\frac{\pi}{3})i}=-c_b;\nonumber
\end{eqnarray}
with $0\le \theta<\pi$ in the complex plane. In fact, there is only a relative meaning between $r, g$ and $b$. The modulus of each color charge is 1, so it is also called the unit color charge. We can also represent them in the form of unit vectors in the unit circle, as shown in FIG.\ref{vecc}.

Obviously, they satisfy
\begin{eqnarray}
  \vec{c}_i+\vec{c}_{\bar{i}} &=& 0 , \label{mcolor} \\
  \vec{c}_r+\vec{c}_g+\vec{c}_b &=& 0, \label{Bcolor}
\end{eqnarray}
with $i=r,g,b$, ensuring the color neutrality of mesons and baryons. Color charges are quantized, which means that all color charges can only be integer mutiples of the three unit color charges aforementioned. The color charges can be added together as
\begin{equation}
C =\sum_{i=r,g,b} (n_ic_i+n_{\bar{i}}c_{\bar{i}}).\label{colors}
\end{equation}
For example, a di-quark composed of a red color charge and a blue color charge results in an anti-green color charge. This allows the existence of particles with color charges other than unit color charges. The interaction between a pair of stationary color charges in vacuum is  Coulomb-like defined as
\begin{equation}\label{CCF}
  \boldsymbol{F}_{C_1C_2}= Z \frac{ C_1\cdot C_2}{r^3}\boldsymbol{r}.
\end{equation}
Here, $Z$ is called the vacuum color gravitational constant and has units of $Nm ^ 2 $.  The dot product of two color charges is defined as the dot product of the vectors represented by the two color charges. So, the dot product $c_{ij}=\vec{c}_i\cdot \vec{c}_j$ between two unit color charges is the matrix element of the following matrix,
\begin{equation}\label{CDC}
  CC=Q\left[
      \begin{array}{ccc}
        1 & -1/2 & -1/2 \\
        1/2 & 1 & -1/2 \\
        1/2 & 1/2 & 1 \\
      \end{array}
    \right]
\end{equation}
with $Q =-1$ for a quark and an anti-quark, and otherwise $Q =1$. This definition is similar to the one in Ref. \cite{ctan}.
\begin{figure}[h!]
\center{\includegraphics[width=0.4\textwidth]{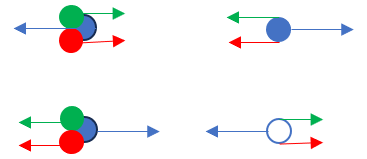}}
\caption{The interaction between a blue quark(antiquark) and a three-quark cluster.}\label{figqH}
\end{figure}
As shown in FIG.\ref{figqH}, one can verify that the total interaction between a blue quark (or antiquark) and a three quark cluster (with a total color charge of zero) is zero:
\begin{equation}
F=\sum_{i}^{r,g,b}Z\frac{\pm c_b \cdot c_i}{r^2}=Z\frac{\sum_i^{r,g,b}\pm c_b\cdot c_i}{r^2}=0.
\end{equation}
Obviously, when taking infinity as the zero point of color potential energy, the color potential energy between two color charges can be calculated by
\begin{equation}\label{CSE}
 E_p=\int_r^\infty\frac{F_{C_1C_2}}{r^2}dr=-Z\frac{C_1\cdot C_2}{r}.
\end{equation}
When $C_1=-C_2$ and $|C_1|=1$, the result above corresponds to the third term in Eq. (\ref{pota})-- the Coulomb-like term.

\section{The color magnetic field from the motion of color charge}\label{s3}
\begin{figure}[h!]
\center{\includegraphics[width=0.4\textwidth]{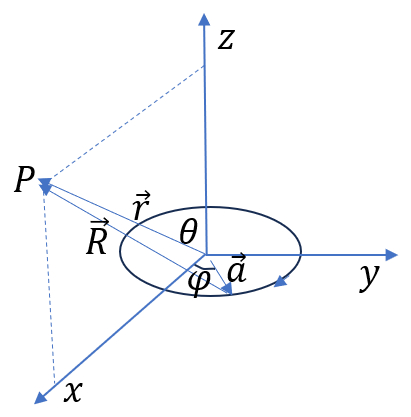}}
\caption{Calculation of color magnetic field generated by a circular color flow.}\label{figCB}
\end{figure}

The collective motion of color charges generates color flow. Similar to the electrical current intensity, the current intensity of color flow is defined as the amount of color charge flowing through a cross-section per unit time:
\begin{equation}\label{Ic}
  I_c=\frac{|\Delta C|}{\Delta t}=\frac{|\sum_{i=r}^b (n_i-n_{\bar{i}})c_i|}{\Delta t}.
\end{equation}
It is assumed that color flow can generate a color magnetic field, modeled after Biot Savart's law:
\begin{equation}\label{Bc}
  \boldsymbol{B}_c=\int_l T\frac{I_cd\boldsymbol{l}\times \boldsymbol{R}}{R^3}.
\end{equation}
Here, $T$ is a parameter under vacuum, and its value needs to be measured directly or indirectly through experiments.
For example, consider the color magnetic field generated by a circular color flow. As shown in FIG. \ref{figCB}, assuming the radius of the circular color current is $a$ and the color current intensity is $I_c$, the following formula can be derived by simulating the magnetic field generated by a circular current \cite{c15}

\begin{eqnarray}
  B_{cx} &=& TI_c ar\cos\theta\int_0^{2\pi}\frac{\cos\varphi }{(r^2+a^2-2ra\sin\theta\cos\varphi)^{3/2}}{d\varphi},\nonumber\\
  B_{cy} &=& 0, \nonumber\\
  B_{cz} &=& TI_c\int_0^{2\pi}\frac{a^2-ar\sin\theta\cos\varphi}{(r^2+a^2-2ra\sin\theta\cos\varphi)^{3/2}}{d\varphi}.\label{eqBc}
\end{eqnarray}

For points on the color flow plane, $\theta=\pi/2, \sin\theta=1$, and $\cos\theta=0$, therefore, $B_{cx}=B_{cy}=0$, and
\begin{eqnarray}
B_{cz} &=& TI_c\int_0^{2\pi}\frac{a^2-ar\cos\varphi}{(r^2+a^2-2ra\cos\varphi)^{3/2}}{d\varphi} \nonumber\\
&=&2TI_c\left[\frac{1}{a-r}E(k)+\frac{1}{a+r}K(k)\right] \nonumber \\
&=&2TI_cX(a,r).\label{eqBcz}
\end{eqnarray}

\begin{figure}[h!]
\center{\includegraphics[width=0.45\textwidth]{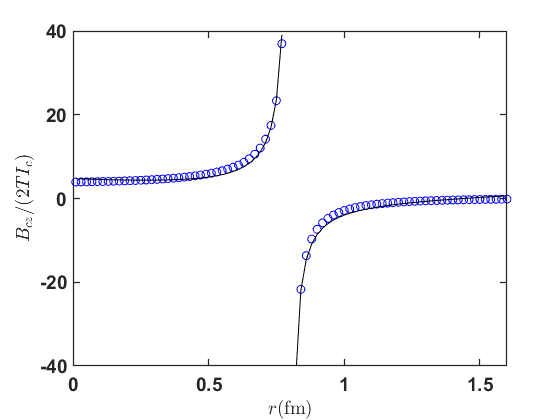}}
\caption{The distribution of the color magnetic field generated by a circular color flow in the color flow plane. The small circles in the figure are calculated according to  Eq. (\ref{eqBcz}), and the solid lines are fitted by Eq. (\ref{mfit})}\label{figBcz}
\end{figure}

Here, $E(k)$ and $K(k)$ are the elliptic functions ellipticE and ellipticK, respectively, with $k=2\sqrt{ar}/(a+r)$. The color magnetic field diverges on the color flow circular line, as shown in FIG. \ref{figBcz}.

\begin{figure}[h!]
\center{\includegraphics[width=0.4\textwidth]{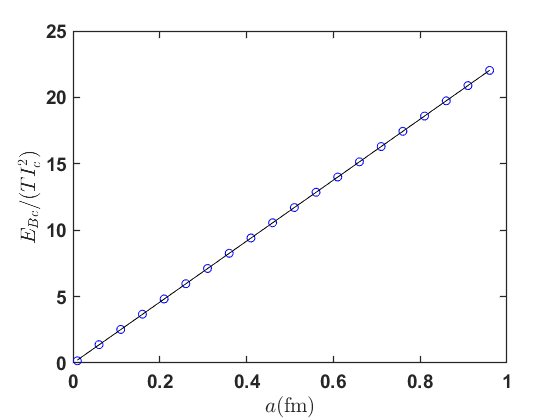}}
\caption{The open circles are the results calculated by Eq. (\ref{eqEBc}). The solid line is the fit by Eq. (\ref{eqEBcf}).}\label{figLc}
\end{figure}

For different radii of circular currents, we have obtained a simple explanatory formula for the inner and outer color magnetic fields of the circle through segmented fitting. Here, $X_i$ and $X_o$ represent the values of $B_{cz}/2TI_c$ inside and outside the circle, respectively.

\begin{eqnarray}
  a=0.5, && X_i=1.2215\frac{1}{a-r}+9.9248(a-r), \nonumber\\
        && X_o=-0.8433\frac{1}{r-a}+5.0440(r-a);\nonumber\\
  a=0.6, && X_i=1.2001\frac{1}{a-r}+7.0576(a-r), \nonumber\\
        && X_o=-0.8593\frac{1}{r-a}+3.6412(r-a); \nonumber\\
  a=0.7, && X_i=1.1833\frac{1}{a-r}+5.2816(a-r),\nonumber\\
        && X_o=-0.8717\frac{1}{r-a}+2.7549(r-a); \nonumber\\
  a=0.8, && X_i=1.1696\frac{1}{a-r}+4.1042(a-r),\nonumber\\
        && X_o=-0.8817\frac{1}{r-a}+2.1587(r-a);\label{mfit} \\
  a=0.9, && X_i=1.1583\frac{1}{a-r}+3.2828(a-r),\nonumber\\
        && X_o=-0.8900\frac{1}{r-a}+1.7381(r-a);\nonumber \\
  a=1.0, && X_i=1.1487\frac{1}{a-r}+2.6866(a-r),\nonumber\\
        && X_o=-0.8970\frac{1}{r-a}+1.4301(r-a).\nonumber
\end{eqnarray}
The color magnetic field energy stored in a color flow ring can be calculated as follows:
\begin{equation}
  E_{Bc} = \frac{1}{2}I_c\int \boldsymbol{B}_c\cdot d\boldsymbol{S}= T I_c^2 \int_0^a X_i 2\pi r dr .\label{eqEBc}
\end{equation}
Since the value of $T$ has not yet been determined, FIG. \ref{figLc} provides the calculated values and fitting relationships of $\frac{E_{Bc}}{TI_c^2}$ versus $a$, using the formula Eq. (\ref{eqEBcf})

\begin{equation}\label{eqEBcf}
  \frac{E_{Bc}}{TI_c^2}=22.97a.
\end{equation}
Now we can apply our scenario to mesons. When one quark orbits another quark in a circular motion with a radius $r$, its equivalent color flow intensity is
\begin{equation}
I_c=|c|\frac{v}{2\pi r}=\frac{v}{2\pi r}.
\end{equation}
Taking into account the centripetal force provided by the color charge force,
\begin{equation}
Z\frac{1}{r^2}=m\frac{v^2}{r},\label{XXL}
\end{equation}
the color magnetic energy stored in the meson is\begin{equation}
  E_{Bc}=22.97TI_c^2r=0.5818T\frac{Z}{mr^2}.
\end{equation}
This is the fourth term in the potential energy Eq. (\ref{pota}), which is inversely proportional to the square of the distance.

\section{the harmonic oscillator potential originated from spin}\label{s4}
\begin{figure}[h!]
\center{\includegraphics[width=0.4\textwidth]{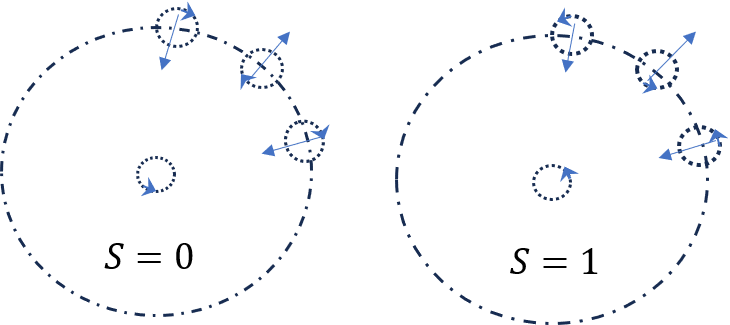}}
\caption{Schematic diagram of spin interaction. When the color flow is parallel, they are attracted to each other. When it is antiparallel, they are repelled.}\label{figSS}
\end{figure}

In quantum mechanics, spin is an intrinsic property of particles. However, in classical terms, we propose that particle spin is an external manifestation of its internal structure. Here, we assume that the quark color charge undergoes circular motion around its own central axis, equivalent to a circular color flow ring, thereby possessing a colored magnetic moment. For a meson system composed of a pair of positive and negative quarks, as shown in FIG. \ref{figSS},  the color magnetic moments of the two quarks must be coplanar due to the effect of the chromomagnetic torque. Consequently, their spin orientations can only have two states: parallel or antiparallel. During their respective rotations, when the directions of the two color flows are parallel, the spin interaction is attractive. When the directions are antiparallel, the spin interaction is repulsive. If the color flow directions are perpendicular to each other, no force acts between them.

Therefore,  spin induced interactions can be described by a harmonic oscillator, whose dynamic equilibrium position lies along the circumference of quark's orbital motion:
\begin{equation}F_{S_1S_2}=-k(r'-r)=-\frac{1}{2}m\omega^2(r'-r).\end{equation}
Here, $k$ is the elastic coefficient, and $\omega$ is the angular frequency. The total energy of the oscillator is
\begin{equation}E_s=\frac{1}{2}kA^2=\frac{1}{2}k(r_m-r)^2.\end{equation}
Expanding the right hand side of the above equation yields the remaining three terms of Eq. (\ref{pota}). Due to the extremely small size of $r_m-r$ (comparable to the radius of quarks), the vibrational energy is expressed using the quantum mechanical harmonic oscillator energy formula:
\begin{equation}
E_s=(L+\frac{1}{2})\hbar \omega_{nL}, \quad (L=0,1,2,\cdots,n-1 ).\label{XZZ}
\end{equation}

\begin{figure}[h!]
\center{\includegraphics[width=0.4\textwidth]{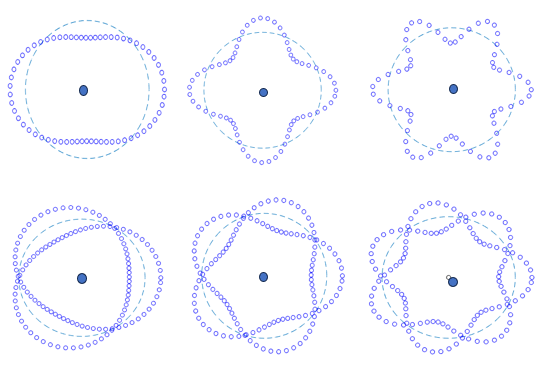}}
\caption{Meson mechanics structure diagram with $n=1,2,3$ from left to right. The top row corresponds to $S=0$, while the bottom row is for $S=1$.}\label{fig1}
\end{figure}

The top row in FIG. \ref{fig1} represents the state where two quarks have the parallel spins, while the bottom row represents the state with antiparallel spins, i.e. $S_1+S_2=0,1 $. Clearly, the stationary orbital motion period must satisfy a specific relationship with the spin period.  As shown in FIG. \ref{fig1}, for a meson system to be in a stable state, the period of quark circular motion $T_{\theta}=2\pi r/v$ must be an odd (for $S=0$) or even(for $S=1$) multiple of the vibration half-period, i.e.,

\begin{equation}
\frac{2\pi r_{nL}}{v_{nL}}=\left\{\begin{array}{ll} (2n-1)\frac{\pi}{\omega_{nL}},& \quad (S=0) \\ \\
2n\frac{\pi}{\omega_{nL}},& \quad (S=1)\end{array}\right.\quad n=1,2,\cdots
\end{equation}
According to Eq. (\ref{XXL}), one can obtain
\begin{equation}\label{ern}
  r_{nL}^3=\left\{\begin{array}{ll} (n-\frac{1}{2})^2\frac{Z}{m\omega_{nL}^2}, &\quad (S=0)\\ \\
  n^2\frac{Z}{m\omega_{nL}^2},  &\quad  (S=1)  \end{array}\right.
\end{equation}
Note that the terms $\omega_{nL}$ here do not represent the angular velocity of orbital motion, i.e., $\omega_{nL}\ne v_{nL}/r_{nL}$.

\section{The values of $Z$ and $T$ and results}\label{s5}
Due to color confinement, it is impossible to directly measure the values of Z and T by measuring the forces between two free quarks or two color currents. Instead, we can only estimate their values using the masses and radii of certain hadrons measured experimentally.

In the center of mass frame of a meson, the mass of the meson can be calculated as \cite{Mma,Mmb}
\begin{equation}M_{nL}=M_Q+M_{\bar{q}}+E_{nL}.\end{equation}
Here, $E_{nL}$ includes the previously mentioned $E_p,E_{Bc},E_s$, and the kinetic energy $E_k=\frac{1}{2}mv^2=\frac{Z}{2r}$, where the reduced mass is given by
\begin{equation}m=\frac{M_QM_{\bar{q}}}{M_Q+M_{\bar{q}}}.\end{equation}

Therefore, for a meson in a state with quantum number $n$ and $L$, the classical calculation of its mass can be expressed as
\begin{eqnarray}
  M_{nL}&=&M_Q+M_{\bar{q}}-\frac{Z}{2r_{nL}}+0.5818T\frac{Z}{mr_{nL}^2}\nonumber \\
  &+&(L+\frac{1}{2})(n-\frac{1}{2}\delta_{S0})\sqrt{\frac{Z}{mr^3_{nL}}}. \label{erM}
\end{eqnarray}

For the ground state radius $r_{10}$ and mass $M_{10}$ , we have
\begin{equation}
  r_{10}^3=\frac{Z}{4m\omega_{10}^2}\;(S=0)\;\mbox{or}\; \frac{Z}{m\omega_{10}^2}\; (S=1), \label{pir}
  \end{equation}
  \begin{equation}
  M_{10}=M-\frac{Z}{2r_{10}}+0.5818T\frac{Z}{mr_{10}^2}+\frac{1}{2}\sqrt{\frac{\frac{1}{4}(1)Z}{mr_{10}^3}}. \label{pim}
\end{equation}
 In Eq.(\ref{pim}) , the coefficients $1/4$ and $1$  under the square root correspond to the cases of spin 0 and spin 1, respectively. The values of $Z$ and $T$ are universal. This allows us to use a small amount of experimental data to deduce their values. Then, we use the obtained values of Z and T to calculate the results for other mesons and test our model by  the experimental data.

It is important to note that the previous discussion did not consider relativistic effects, which may require corrections for light meson systems. For heavy mesons, relativistic effects are likely less significant, so we use heavy meson data to calculate $Z$ and $T$.

Unfortunately, due to current experimental limitations, directly measuring meson radii is very challenging. However, model calculations of meson sizes have garnered significant interest among scientists. Currently, two important radii are used to describe the size of meson systems: the so-called root mean square (r.m.s.) radius $\langle r^2_{rms}\rangle$ \cite{rms1,rms2} and the charge radius $\langle r^2_E\rangle$ \cite{rch1}. Their definitions are as follows:\begin{eqnarray}
\langle r^2_{rms}\rangle&=&\int_0^\infty r^2[\psi(r)]^2 dr,\\
\langle r^2_E\rangle&=&-6\frac{d^2}{dQ^2}eF(Q^2)|_{Q^2=0},
\end{eqnarray}
where $\psi(r)$ is the radial wave function and  $F(Q^2)$ is the form factor of the meson \cite{FQ2,DHT}. According to calculations in Refs. \cite{c20,Das}, the root mean square (r.m.s.) radius of the $\Upsilon(1S)$ state is approximately $0.2671$ fm. Ref. \cite{c21}, through a comprehensive contact interaction analysis, determined that the ground state charge radius of the pseudoscalar meson $\eta_b$ is about $0.07$ fm.

Now, we use the data of these two mesons to determine the parameters $Z$ and $T$. The mass data of these mesons are obtained from the Particle Data Group (PDG), while the mass of the constituent $b$ quark is taken as $m_b=4.95$ GeV, consistent with Refs. \cite{c20,Das}. These data are listed in TABLE  \ref{tabm}. The obtained values of $Z $ and $T $ are
\begin{table}[!h]
  \centering
\begin{tabular}{|c|c|c|c|c|}
  \hline
   $n^{2S+1}L_j$&Name&$q\bar{q'}$&$\sqrt{\langle r_1\rangle ^2}$ (fm)&$M$(GeV)\\\hline
$1^1S_0$&$\eta_b(1S)$&$b\bar{b}$&0.070 &9.3987  \\
$1^3S_1$&$\Upsilon(1S)$&$b\bar{b}$&0.268&9.4604\\
\hline
\end{tabular} \caption{Meson data taken from PDG \cite{pdg} and Refs \cite{c20,c21}.\label{tabm}}
\end{table}
\begin{equation}
  Z=2.33, \quad T=0.375.
  \end{equation}
In the above discussions, we have adopted the natural units commonly used in high-energy physics, where
$\hbar=c=1$, with GeV as the basic unit. In the International System of Units (SI), $c=2.998\times 10^8$ m/s, $\hbar c\approx 0.197$ GeV$\cdot$ fm. According to Eqs. (8) and (18), the units of
$Z$ and T should be $[Z]=ML^3T^{-2}$ and $[T]=ML$ , respectively. Thus
\begin{eqnarray}
  Z &=& 2.33\times \frac{GeV}{c^2}\times (0.197fm)^3\times \frac{c^2}{(0.197fm)^2} \nonumber \\
  &=& 7.35\times 10^{-26} Nm^2 \nonumber \\
  T&=& 0.375\times \frac{GeV}{c^2}\times (0.197fm)\nonumber \\
  &=&1.32\times 10^{-43} Ns^2
\end{eqnarray}

Comparing  the magnitude of the gravitational forces between a pair of quark and antiquark that are 1 fm apart (with mass and charge of $m_b=m_{\bar{b}}=4.95$ GeV, $q_b=-q_{\bar{b}}=-1/3e$)
\begin{eqnarray}
F_m&=&G\frac{m_bm_{\bar{b}}}{r^2}=5.16\times 10^{-33}N,\\
 F_e&=&k\frac{q_bq_{\bar{b}}}{r^2}=2.56\times 10 N,\\
 F_c&=&Z\frac{c_ic_{\bar{i}}}{r^2}=2.01\times 10^4 N,
 \end{eqnarray}
it is found that the strong interaction, based on color charge, is much greater than the other two, consistent with the hiearchy of force magnitudes known. We use the meson masses provided by the Particle Data Group (PDG) \cite{pdg} as inputs. Applying Eq. (\ref{erM}), we can calculate the corresponding meson radii and compare them with the results from other models, as shown in TABLE \ref{tabm2}
\begin{table}[!h]
  \centering
\begin{tabular}{lclcccl}
  \hline
name           &$q\bar{q'}$&state&{$\tiny M(GeV)$}&$\omega$&$r_{nL}(fm)$&$\sqrt{r^2}$[Ref.] \\\hline
$\eta_b(1S)$   &$b\bar{b}$       &$1^1S_0$      & 9.3987& 0.9093 &0.0700& 0.07\cite{c21}\\
$\Upsilon(1S)$ &$b\bar{b}$       &$1^3S_1$      & 9.4604& 0.8352 &0.2680& 0.2671\cite{c21}\\
$\chi_{b0}(1P)$&$b\bar{b}$       &$1^3P_0$      & 9.8594& 0.2972 &0.4340& 0.39\cite{c21} \\
$\chi_{b1}(1P)$&$b\bar{b}$       &$1^3P_1$      & 9.8928& 0.3586 &0.3830& \\
$h_b(1P)$      &$b\bar{b}$       &$1^1P_1$      & 9.8993& 0.3694 &0.3755& \\
$\chi_{b2}(1P)$&$b\bar{b}$       &$1^3P_2$      & 9.9122& 0.3910 &0.3615& \\
$\eta_b(2S)$   &$b\bar{b}$       &$2^1S_0$      & 9.9987&2.8256 &0.1268& \\
$\Upsilon(2S)$ &$b\bar{b}$       &$2^3S_1$      &10.0233& 2.3623 &0.1730& \\
$\Upsilon_2(1D)$&$b\bar{b}$      &$1^3D_2$      &10.1637& 0.3019 &0.4295&\\
$\chi_{b0}(2P)$&$b\bar{b}$       &$2^3P_0$      &10.2325& 0.5209 &0.4740&\\
$\chi_{b1}(2P)$&$b\bar{b}$       &$2^3P_1$      &10.2555& 0.5448 &0.4600& \\
$h_b(2P)$      &$b\bar{b}$       &$2^1P_1$      &10.2598& 0.6445 &0.3395& \\
$\chi_{b2}(2P)$&$b\bar{b}$       &$2^3P_2$      &10.2687& 0.5575 &0.4530& \\
$\eta_b(3S)$   &$b\bar{b}$       &$3^1S_0$      &10.3268& 2.9857 &0.1718& \\
$\Upsilon(3S)$&$b\bar{b}$        &$3^3S_1$      &10.3552& 2.7877 &0.2030& \\
$\chi_{b1}(3P)$&$b\bar{b}$       &$3^3P_1$      &10.5134& 0.6839 &0.5180& \\
$\chi_{b2}(3P)$&$b\bar{b}$       &$3^3P_2$      &10.5240& 0.6939 &0.5130& \\
$\eta_b(4S)$   &$b\bar{b}$       &$4^1S_0$      &10.5397&3.1094 &0.2092& \\
$\Upsilon(4S)$&$b\bar{b}$        &$4^3S_1$      &10.5794& 3.0422 &0.2320& \\
$\eta_b(5S)$   &$b\bar{b}$       &$5^1S_0$      &10.8202& 3.5592 &0.2261& \\
$\Upsilon(5S)$&$b\bar{b}$        &$5^3S_1$      &10.8761& 3.5696 &0.2420& \\\hline
$\eta_c(1S)$  &$c\bar{c}$        &$1^1S_0$       &2.9839& 0.7948 &0.2090&0.20\cite{c21} \\
$J/\psi(1S)$  &$c\bar{c}$        &$1^3S_1$       &3.0969& 0.9675 &0.2910&\makecell[tl]{0.37\cite{EbD}\\ 0.28\cite{c22}} \\
$\chi_{c0}(1P)$&$c\bar{c}$       &$1^3P_0$       &3.4147& 0.4539 &0.4820&0.43\cite{c21}\\
$\chi_{c1}(1P)$&$c\bar{c}$       &$1^3P_1$       &3.5107& 0.5368 &0.4310&\\
$h_c(1P)$      &$c\bar{c}$       &$1^1P_1$       &3.5254& 0.6262 &0.2450&\\
$\chi_{c2}(1P)$&$c\bar{c}$       &$1^3P_2$       &3.5662& 0.5839 &0.4075&\\
$\eta_c(2S)$   &$c\bar{c}$       &$2^1S_0$       &3.6375& 2.1065 &0.2270&0.386\cite{c22}\\
$\psi(2S)$     &$c\bar{c}$       &$2^3S_1$       &3.6861& 2.1773 &0.2690&0.387\cite{c22} \\
$\psi(3770)$   &$c\bar{c}$       &$2^3P_{0,1}$   &3.7737& 0.6569 &0.5980& \\
$\psi_2(3823)$ &$c\bar{c}$       &$2^3P_2$       &3.8237& 0.6967 &0.5750& \\
$\psi_3(3842)$ &$c\bar{c}$       &$2^3P_3$       &3.8427& 0.7715 &0.5670& \\
$\chi_{c1}(3872)$&$c\bar{c}$     &$2^3P_1$       &3.8717& 0.7347 &0.5550& \\
$\chi_{c0}(3915)$&$c\bar{c}$     &$2^1P_0$       &3.9217& 0.8122 &0.4285& \\
$\chi_{c2}(3930)$&$c\bar{c}$     &$2^3P_2$       &3.9225& 0.7741 &0.5360& \\\hline
$B_c^{+}$    &$c\bar{b}$       &$1^1S_0$             &6.2745& 0.0920 &0.7650&\makecell[tl]{$0.38\sim $\\1.09\cite{rch2}}  \\
$B_c^{+}(2S)$&$c\bar{b}$       &$2^1S_0$             &6.8712& 0.2290 &0.1900& 0.17\cite{c21}\\\hline
$B_s^0$        &$s\bar{b}$       &$1^1S_0$             &5.3669& 0.4514 &0.3680& 0.24\cite{c21}\\
$B_s^*$        &$s\bar{b}$       &$1^3S_1$             &5.5154& 0.7485 &0.4170& \\
$B_{s1}(5830)^0$&$s\bar{b}$      &$1^3P_1$             &5.8286& 0.4359 &0.5980& \\
$B^*_{s2}(5840)^0$&$s\bar{b}$    &$1^3P_2$             &5.8399& 0.4448 &0.5900& \\\hline
$D^+$        &$c\bar{d}$   &$1^1S_0$             &5.2793& 2.5089 &0.1370&\makecell[tl]{$0.10\sim$\\ 0.42 \cite{Das2}} \\
$D^0$        &$c\bar{u}$   &$1^1S_0$             &1.8648& 0.3175 &0.5435&\makecell[tl]{$0.14\sim$\\0.55\cite{Das2}} \\
$D_s^+$      &$c\bar{s}$   &$1^1S_0$             &1.9683& 0.3607 &0.4535&\makecell[tl]{$0.10\sim$\\ 0.4\cite{Das2}} \\
$\eta'$      &$s\bar{s}$   &$1^1S_0$             &0.9578& 0.2934 &0.5990& 0.5\cite{Das2}\\
\hline
\end{tabular} \caption{The results of our model for the particles in PDG. The last column contains reference values from the literature.\label{tabm2}}
\end{table}

\begin{figure}[h!]
\center{\includegraphics[width=0.4\textwidth]{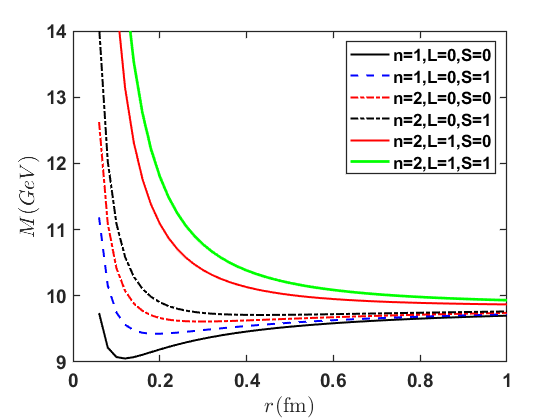}}
\caption{According to Eq. (\ref{erM}), the relationship between the mass and radius of a heavy meson composed of a bottom quark ($b$) and an anti-bottom quark ($\bar{b}$).}\label{fig8}
\end{figure}

From TABLE \ref{tabm2}, one can see that some of results are in good agreement with the results in the literature. It must be noted that existing literature on meson radius calculations employs various models \cite{c17,c18,c20,c21,c22}, each with its own set of parameters, and most focus only on the lowest few states. Our model also relies on data from a few mesons; however, this is due to the current inability to experimentally measure $Z$ and $T$. Once these two values are scientifically determined, a predictable relationship between meson mass and its structural radius can be established. As shown in Fig. \ref{fig8}, we present the mass-radius relationship for several quantum states of $b\bar{b}$ mesons based on Eq. (\ref{erM}), which is of significant importance for understanding hadron structures.  We observe an anomaly in the $^3nS_1$
 state series, particularly in the $\Upsilon(nS)$ mesons. In general, the radius of a meson's excited state is typically larger than that of its ground state. Our analysis suggests that this anomaly is related to spin-dependent terms: for heavy-flavor quarks, the influence of spin energy is significant. As the energy level increases, spin-induced vibrations intensify, causing the parameter $\omega$ to increase. According to Eq. (26), this leads to an increase in $v$. When the angular momentum $L$ remains constant, this leads to a reduction in radius $r$. This is left for the experimental examination in the future.

\section{Summary} \label{s6}
We propose a numerical representation method for the color properties of quarks, facilitating the convenient superposition of color charges and predicting the existence of multi-color-charge states. In analogy with classical electromagnetic field theory, we introduce novel concepts such as ``color force", ``color current", and ``color magnetic field". We further explore the structure of mesons composed of quark-antiquark pairs using Bohr model of hydrogen atom. Using masses of $\eta_b(1S)$  and $\Upsilon(1S)$ from the Particle Data Group (PDG) and their radii from literature, we estimate the fundamental parameters $Z$  for color charge interactions and $T$ for color current interactions in our model. With the determined values, we calculate the masses and radii for several other mesons and compare the results with available literature, finding close agreements.  We admit that more precise calculations of these physical quantities can be achieved using quantum chromodynamics (QCD) theory. We also expect that the classical description of inter-quark interactions in this paper may give a simple and intuitively picture worthy of discussion.

\section*{Acknowledgments}
This work was supported in part by the Department of Education (grant No. 21A0541) and Natural Science Foundation (grant No.2025JJ50382) of Hunan province, China.

\end{document}